\documentclass[aps,pra,reprint,showpacs,floatfix]{revtex4-1}

\usepackage{amsmath}
\usepackage{txfonts}

\usepackage{microtype}
\usepackage{graphicx}
\usepackage{color}
\usepackage{ulem}

\graphicspath{{figures/}}

\begin{document}
\title{Tunneling dynamics in multiphoton ionization and attoclock calibration }

\author{Michael \surname{Klaiber}}
\author{Karen Z. \surname{Hatsagortsyan}} \thanks{\mbox{Corresponding author: k.hatsagortsyan@mpi-k.de }}
\author{Christoph H. \surname{Keitel}} 
\affiliation{Max-Planck-Institut f\"ur Kernphysik, Saupfercheckweg 1, 69117 Heidelberg, Germany}

\date{\today}

\begin{abstract}

The intermediate domain of strong-field ionization between the tunneling and the multiphoton regimes is investigated using the strong field approximation and the imaginary-time method. An intuitive model for the dynamics is developed which describes the ionization process within a nonadiabatic tunneling picture with a coordinate dependent electron energy during the under-the-barrier motion. The nonadiabatic effects in the elliptically polarized laser field induce a transversal momentum shift of the tunneled electron wave packet at the tunnel exit, a delayed appearance in the continuum as well as a shift of the tunneling exit towards the ionic core. The latter significantly modifies the Coulomb focusing during the electron excursion in the laser field after exiting the ionization tunnel. We show that nonadiabatic effects are especially large when the Coulomb field of the ionic core is taken into account during the under-the-barrier motion. The simpleman model modified with these nonadiabatic corrections provides an intuitive background for exact theories and 
%can explain the recent experimental results of Boge et al., Phys. Rev. Lett. 111, 103003 (2013) aimed at probing nonadiabatic effects in tunnel ionization. Our findings also 
has direct implications for the calibration of the attoclock technique which is used for the measurement  of the tunneling delay time.

\end{abstract}


\maketitle

\begin{figure} 
    \begin{center}
      \includegraphics[width=0.4\textwidth]{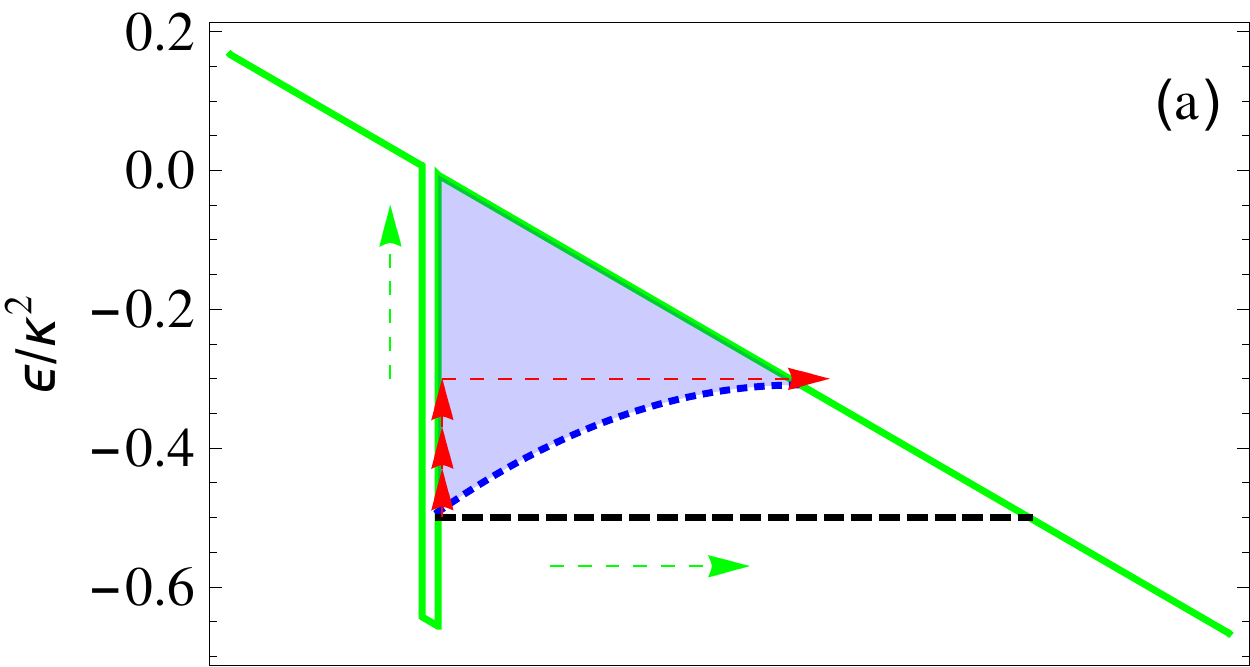}
      \includegraphics[width=0.4\textwidth]{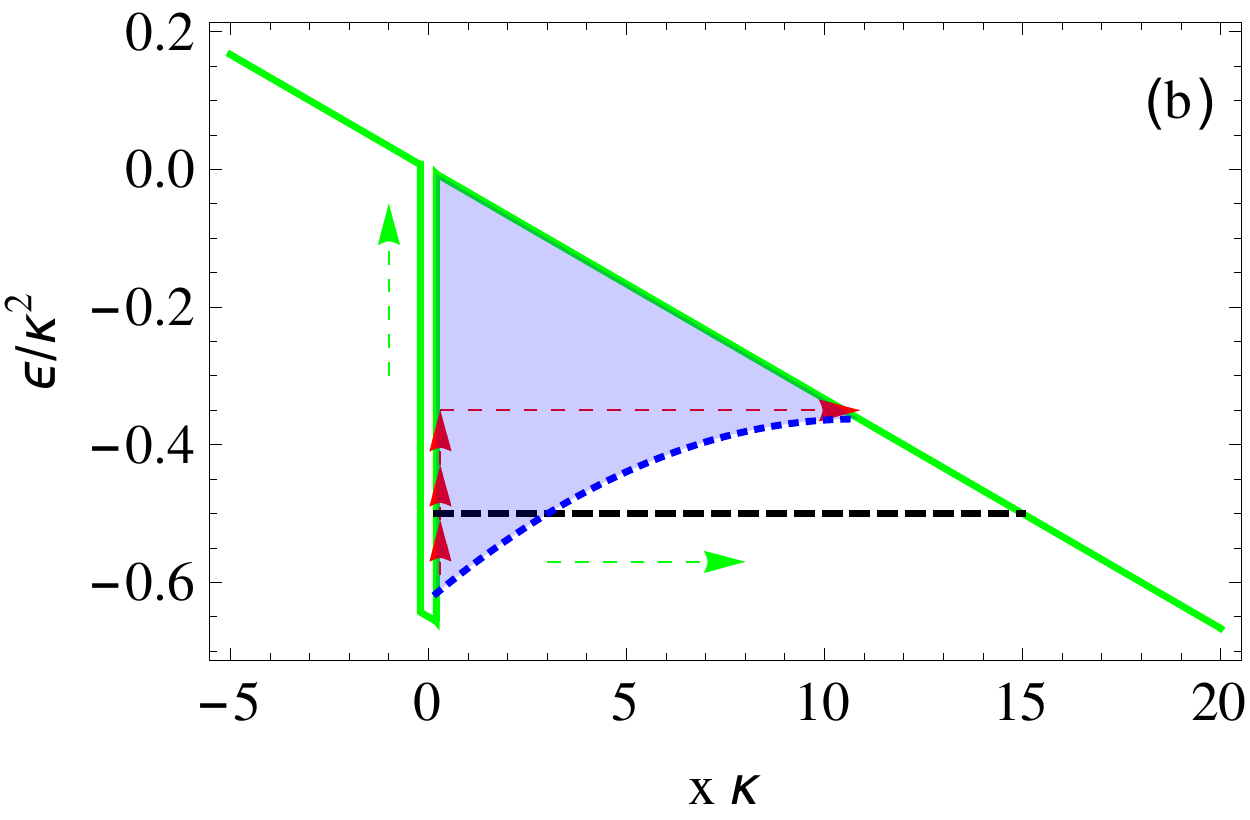}
      \caption{(color online) The tunneling barrier of ionization in the case of a short-range atomic potential (solid, green line). The electron energy during the under-the-barrier motion: nonadiabatic (short-dashed, blue) and adiabatic (quasi-static) picture (long-dashed, black) in a (a) linearly or (b) circularly polarized laser field. The horizontal channel (tunneling) and vertical channel (multiphoton ionization) are shown schematically by arrows. The interpretation of  nonadiabatic tunneling as absorption of photons followed by tunneling with higher energy is shown with the red pathway in (a).} 
 \label{picturelp}
    \end{center}
  \end{figure}

In intense near-infrared laser fields, when the photon energy is much less than the ionization energy of the atomic system, the atomic ionization happens via multiphoton processes \cite{Protopapas_1997,Becker_2002}. 
The multiphoton and tunneling regimes have been identified as well-known asymptotic limits~\cite{Keldysh_1965} with the latter following an especially intuitive tunneling picture. In this case the laser field is so strong that the bound electron tunnels with a constant energy through the (quasi-)static potential barrier formed by the laser field and the atomic potential  (horizontal channel in phase-space at a constant energy of the ionizing electron \cite{Ivanov_2005,Boge_2013}, see Fig. \ref{picturelp}). The quasi-static (adiabatic) dynamics is characterized by an asymptotically small Keldysh-parameter $\gamma\ll 1$, where $\gamma=\kappa\omega/E_0$,  $I_p=\kappa^2/2$ is the ionization potential,  $E_0$ the laser electric field strength, and $\omega$ the laser angular frequency. In the opposite asymptotic limit  $\gamma\gg 1$ of the multiphoton regime, the electron release from the bound state happens at the atomic core via overcoming the atomic potential due to the absorption of multiple laser photons by the bound electron (vertical channel in phase-space at a constant coordinate of the ionizing electron \cite{Ivanov_2005,Boge_2013}, see Fig. \ref{picturelp}). The strong field ionization in both regimes can be described analytically in the strong field approximation \cite{Keldysh_1965,Faisal_1973,Reiss_1980} and the imaginary time method \cite{Perelomov_1966a,Perelomov_1966b,Perelomov_1967a,Popov_1967,Popov_2004u}, which is applied also for arbitrary Keldysh parameters \cite{Yudin_2001b,Mur_2001,Popruzhenko_2008c,Bondar_2008,Barth_2011}. It is straightforward to deduce from the quasi-static theory the parameters of the tunneling  picture, such as the coordinate of the tunnel exit and the electron  momentum at the tunnel exit \cite{Pfeiffer_2012}. However, it is not clear how the intuitive picture is gradually transformed from the horizontal tunneling to the vertical multiphoton channel within the intermediate regime. While the intuitive picture is appealing per se, it allows also to predict how the tunneling exit coordinate and the electron momentum at the  exit are qualitatively modified in the nonadiabatic domain. The latter is important because these parameters are required for the attoclock calibration, which in a recent series of experiments is employed for measuring the tunneling time delay of ionization  \cite{Eckle_2008a,Eckle_2008b,Pfeiffer_2012,Pfeiffer_2013}. In the attoclock \cite{Eckle_2008a} the time of the electron's appearance in the continuum is mapped onto the angle of the photoelectron emission. For its calibration the emission angle should be corrected on the amount originated from  Coulomb focusing which is determined by the tunneling parameters.

Recent experimental investigations  of nonadiabatic effects for the attoclock calibration indicated no significant impact of these effects on the distribution of the photoelectron momentum up to a Keldysh parameter of $\gamma\approx 3.8$ \cite{Boge_2013} and the difference  between the quasi-static calculations and experimental results was attributed to a tunneling delay time. However, numerical simulations \cite{Ivanov_2014} and a R-matrix theory calculation \cite{Kaushal_2013} concluded that the observed photoelectron emission momentum distribution are explainable with a vanishing tunneling time delay when the Coulomb field of the atomic core is fully taken into account \cite{Torlina_2014}.

In this letter we put foward an intuitive picture for the intermediate regime of ionization describing it as tunneling through a classically forbidden region with a coordinate dependent rising energy  due to the time-dependent barrier. The picture allows to deduce in a simple way the characteristics of the under-the-barrier motion and shows how the semi-classical theory of \cite{Boge_2013} should be remedied to describe the observed photoelectron spectra, explaining the discrepancy between results of \cite{Boge_2013} and \cite{Torlina_2014}. Nonadiabatic effects induce a transversal momentum shift of the electron at the tunneling exit, a delayed appearance in the continuum  as well as a shift of the tunneling exit coordinate towards the ionic core. While for the asymptotic momentum distribution in the case of a short range atomic potential all three effects almost compensate each other, the effect of the shift of the tunneling exit dominates when the Coulomb field of the atomic core is accurately taken into account in the under-the-barrier motion. This has a decisive impact on the Coulomb focusing during the motion in the continuum after tunneling and, consequently, on the final momentum distribution and on the calibration of the attoclock.

For the analysis we employ the strong-field approximation (SFA) with the saddle point-approximation including nonadiabatic corrections and quantify the effect of nonadiabatic corrections on the attoclock calibration.
For the sake of simplicity let us begin with modelling the atomic system by a three-dimensional short-range potential $V^{(z)}$. The ionization dynamics is described by the  Hamiltonian
\begin{eqnarray}
  H=\hat{\mathbf{p}}^2/2+V^{(z)}(\mathbf{r})+\mathbf{r}\cdot \mathbf{E}(t),
\end{eqnarray}
where $\hat{\mathbf{p}}$ is the momentum operator, $\mathbf{E}(t)=-E_0\left(\cos\omega t,\epsilon\sin\omega t\right)$ the laser field of elliptical polarization with laser ellipticity $\epsilon$ 
(atomic units are used throughout the paper). According to SFA the ionized wave-packet in momentum space at some time $t$, when the laser pulse is turned off, reads \cite{Becker_2002}:
\begin{eqnarray}
  \psi(\mathbf{p},t)=\int^t_{-\infty} dt'\langle\mathbf{p}|V^{(z)}|\phi^{(b)}\rangle\exp[-i S^L(t,t',\mathbf{p})+i I_pt'],
  \label{int}
\end{eqnarray}  
with the constant matrix element of the short-range-potential  $\langle\mathbf{p}|V^{(z)}|\phi^{(b)}\rangle$ and the bound state $|\phi^{(b)}\rangle$. $S^L(t,t',\mathbf{p})=\int^t_{t'}d\tilde{t}\,{\cal E}(\tilde{t},\mathbf{p})$ is the classical  action of the active electron in the laser field,   
$\,{\cal E}(\tilde{t},\mathbf{p})=(\mathbf{p}+\mathbf{A}(\tilde{t}))^2/2$ the energy of the electron in the laser field  with the asymptotic electron momentum $\mathbf{p}$ and the laser vector potential $\mathbf{A}(t)$, $\mathbf{E}(t)=-\partial_t\mathbf{A}(t)$. We assume that the photon energy of the laser field is much smaller than the ionization ($I_p$) and ponderomotive ($U_p=E_0^2/2\omega^2$) energies $\omega\ll I_p,U_p$. Then the integral in Eq.~(\ref{int}) can be solved via the saddle-point method (SPM) which defines the initial time of ionization $t'=t_s$ via ${\cal E}(t_s,\mathbf{p})=-\kappa^2/2$, describing the energy conservation when the electron starts to leave the bound state. This, here called,  saddle time $t_s$ is complex due to the negative binding energy $-\kappa^2/2$.
The motion of the electron can be described by two steps. The first step is a motion in the classically forbidden region where the time runs from the initial complex saddle time to the real time axis. When the time reaches the real axis at $t_e$, representing the tunnel exit time, the free motion in the laser field begins and from that time on runs along the real axis. The ionization probability does not change after $t>t_e$ and, therefore, is determined by the exponent
\begin{eqnarray}
  \Gamma\sim \left|\psi(\mathbf{p},t)\right|^2\sim \exp[-2iS(t_e,t_s,\mathbf{p})],
  \label{Gamma}
\end{eqnarray}
which is a function of the final momentum $\mathbf{p}$ or, equivalently, of the tunneling phase~\cite{Yudin_2001b}. Thus, in the physical situation suitable for the SPM ($\omega\ll I_p,U_p$), at any value of the Keldysh parameter, during the ionization the electron penetrates the classically forbidden region. This dynamics can be termed as tunneling, although at large $\gamma$ the energy is not conserved during the tunneling as we will show below.

\begin{figure}
    \begin{center}
      \includegraphics[width=0.23\textwidth]{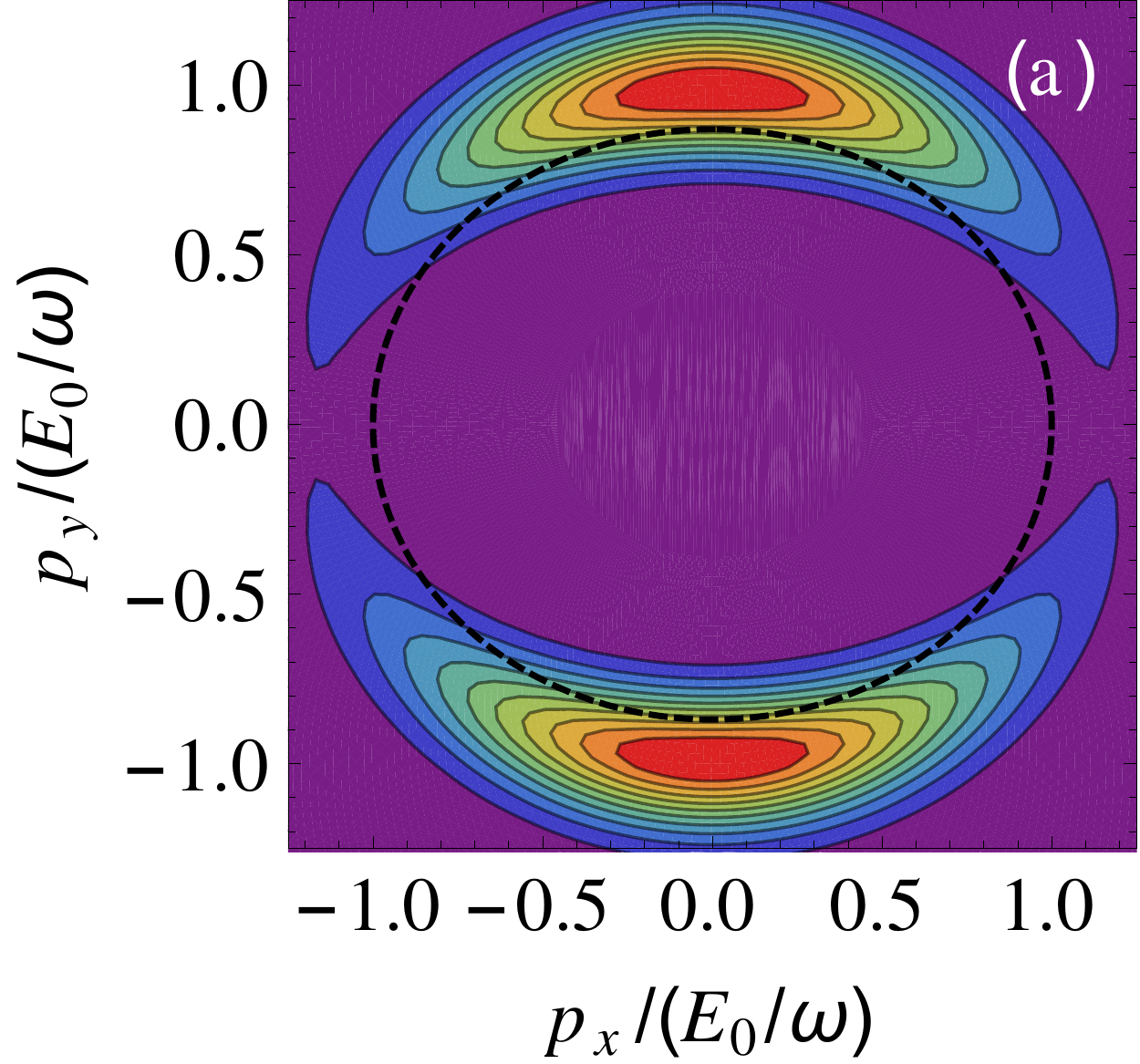} 
      \includegraphics[width=0.23\textwidth]{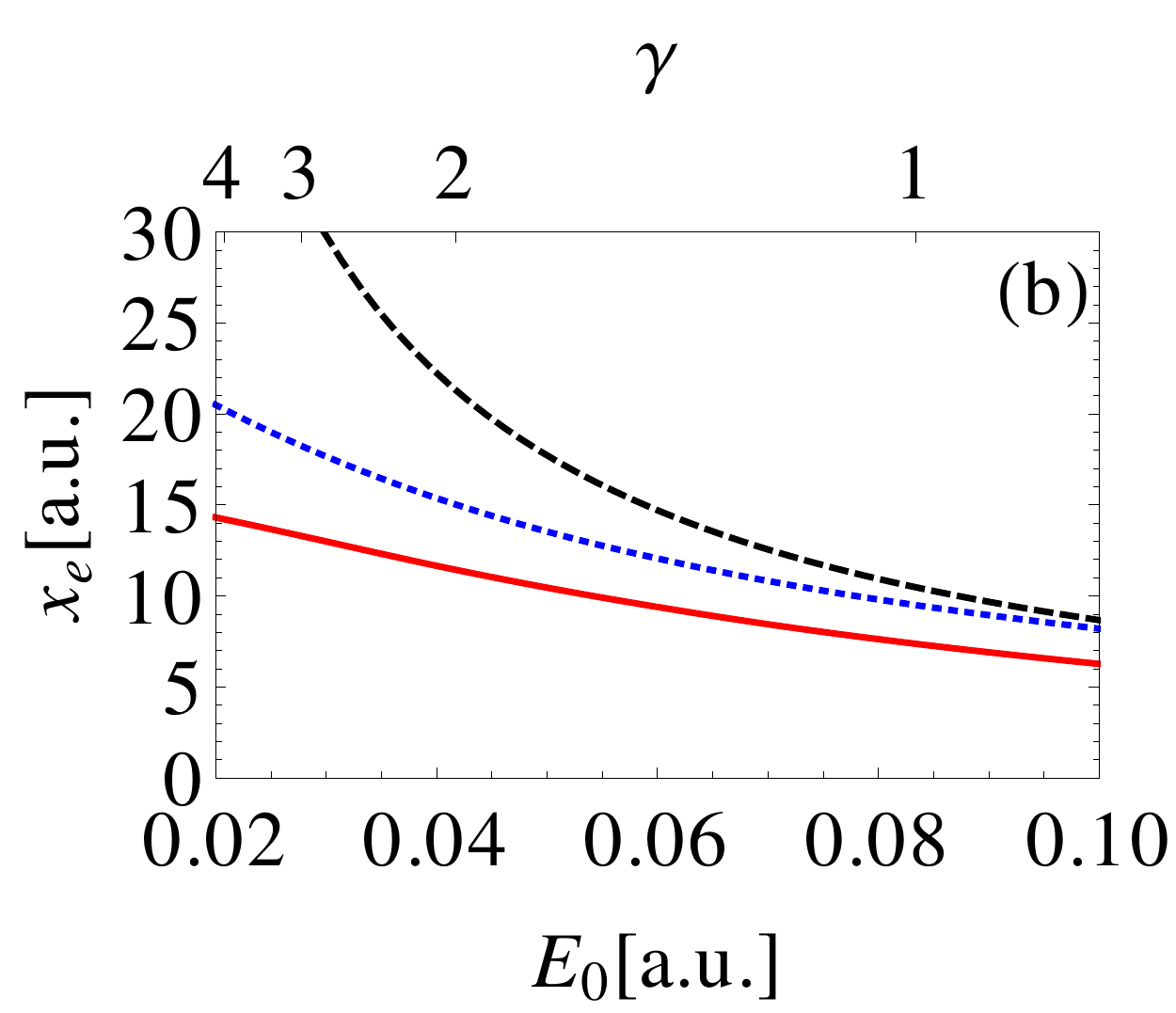}
      \includegraphics[width=0.23\textwidth]{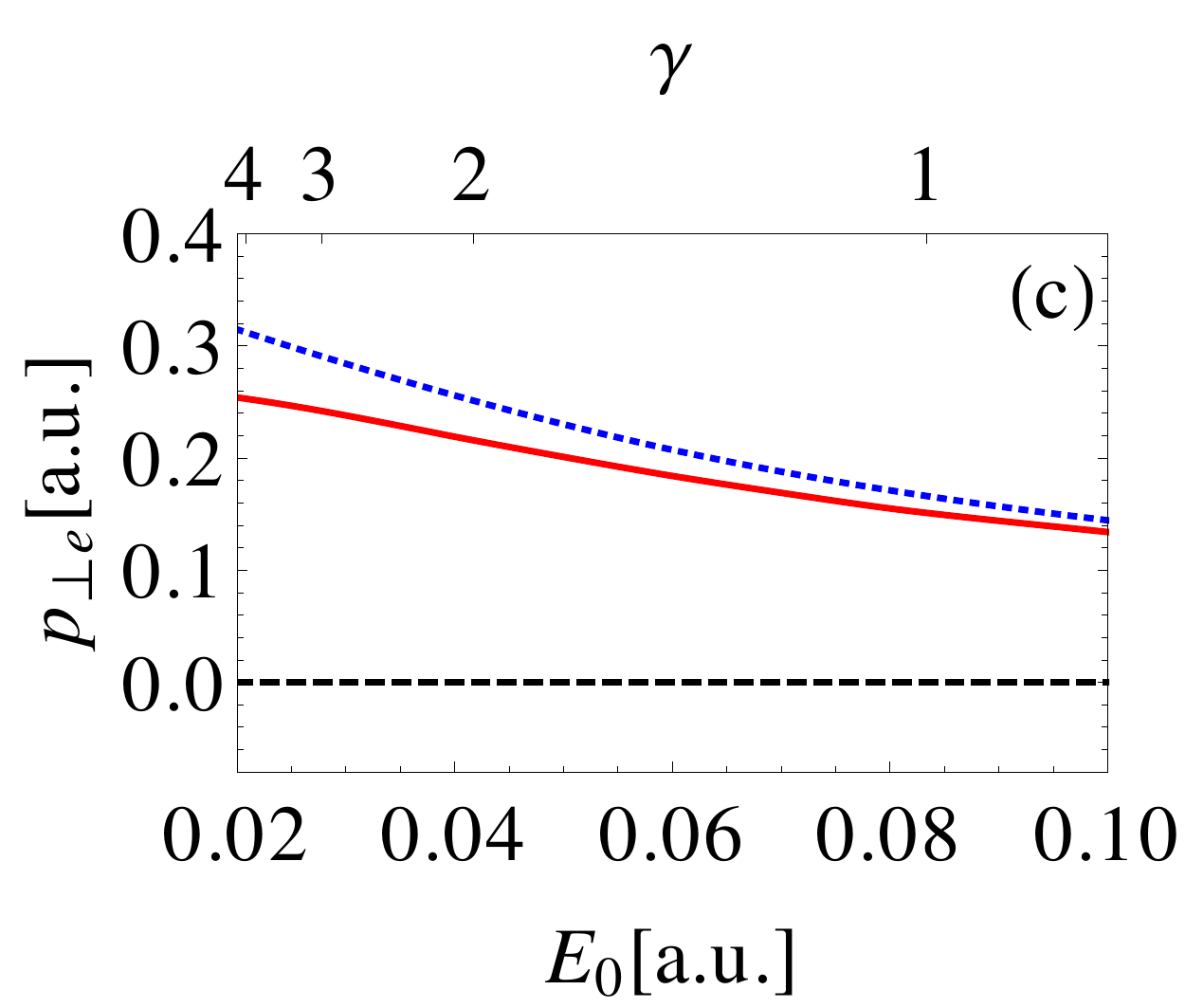} 
      \includegraphics[width=0.23\textwidth]{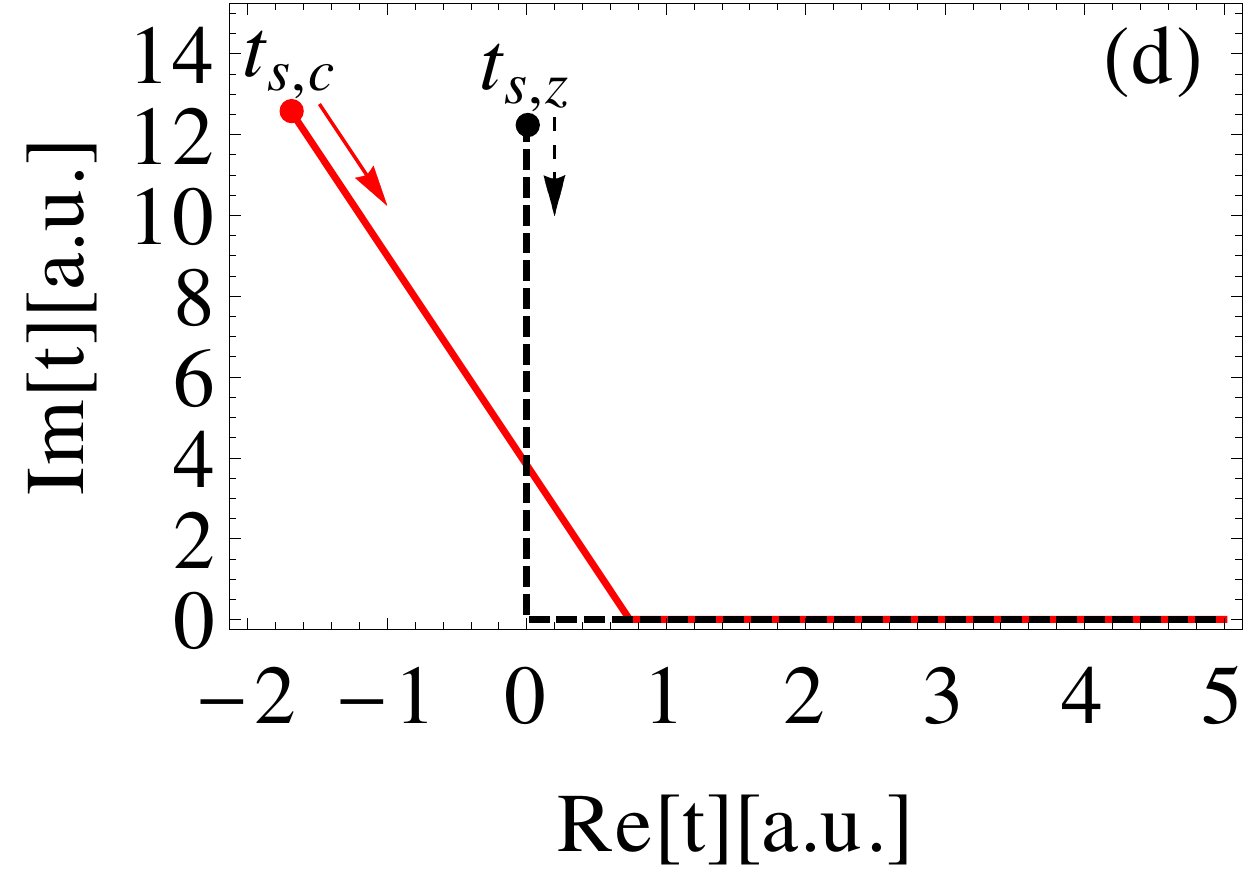}
      \includegraphics[width=0.23\textwidth]{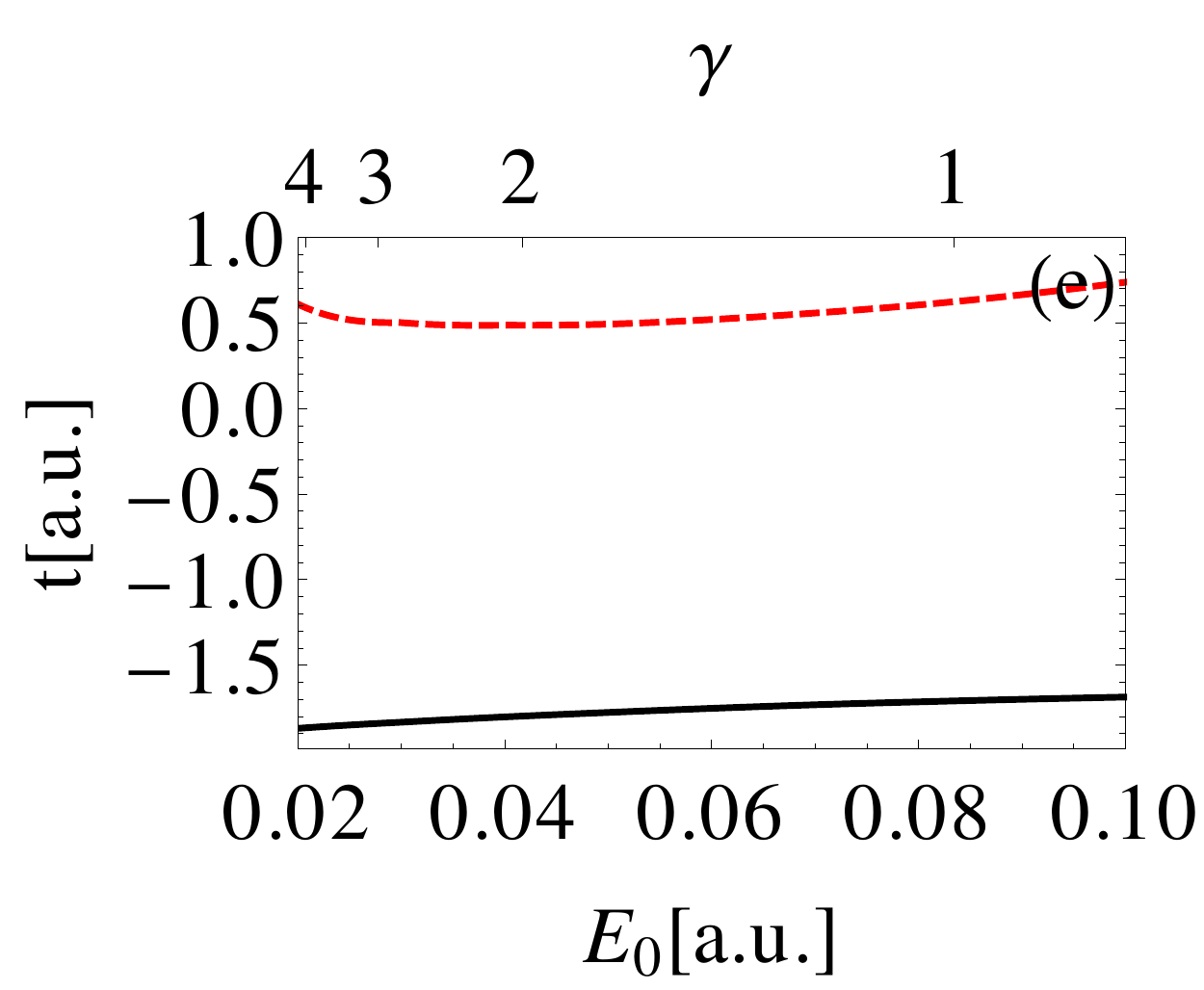}
      \includegraphics[width=0.23\textwidth]{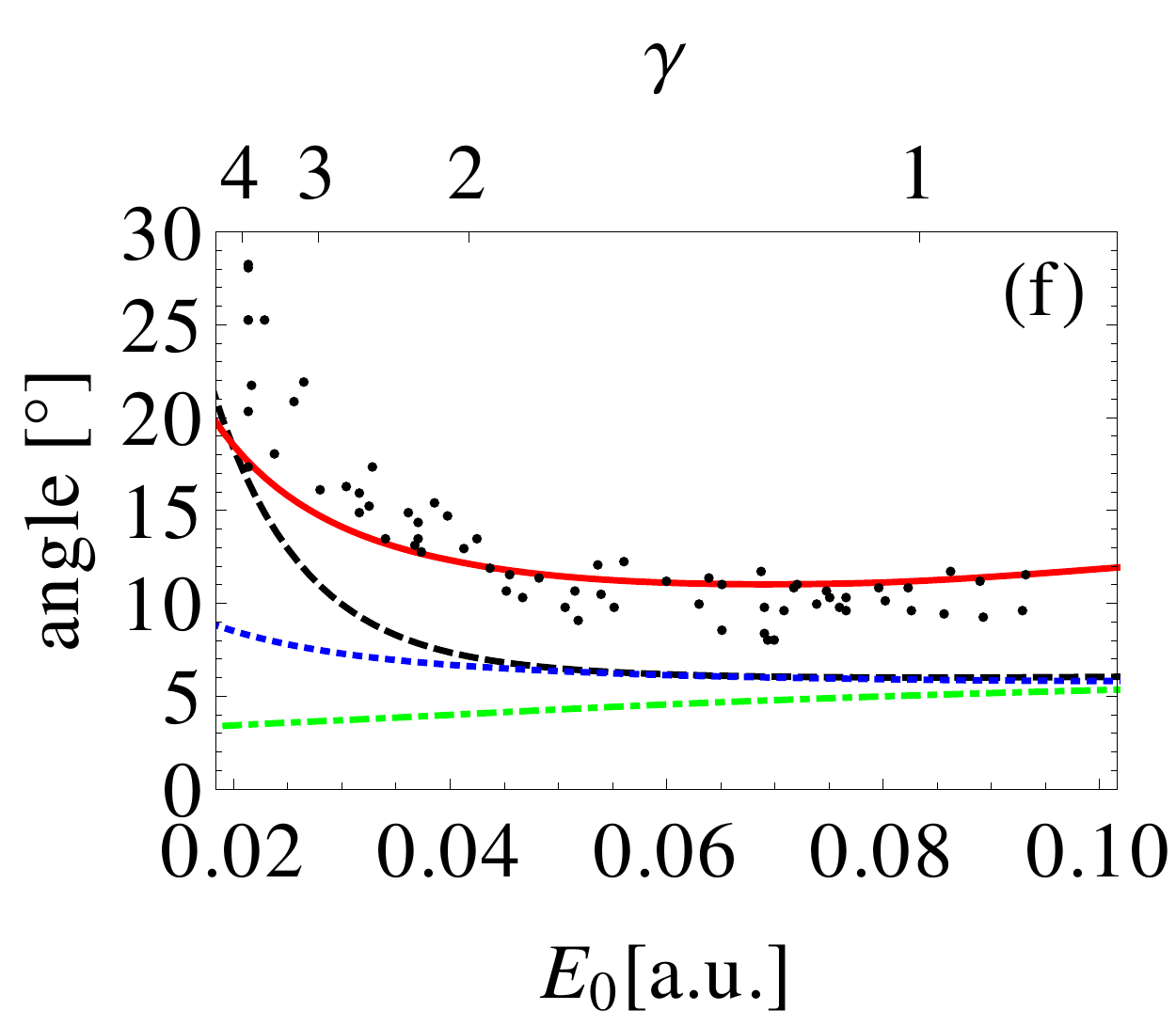}
      \caption{Ionization in a laser field with ellipticity of $\epsilon=0.87$. 
        (a) The asymptotic momentum distribution, and the corresponding quasi-static result shown as a dashed ellipse for $\gamma=1$;
        (b) The exit coordinate vs the Keldysh parameter; (c) The most probable transverse momentum at the tunneling exit $p_{\perp e}$ vs the Keldysh parameter; (d) The complex time contour during tunneling in the Coulomb-potential (red, solid) and the zero-range-potential (black, dashed) for $E_0=0.1$ a.u. (the arrows show the integration direction); (e) The real (black, solid) and imaginary (red, dashed) parts of the complex saddle time vs the field strength $E_0$; (f) The emission angle of the most probable photoelectrons vs the Keldysh parameter. In (b), (c), and (f) (red, solid) the nonadiabatic case for the Coulomb potential, (black, long-dashed) the quasistatic case in the Coulomb potential, and (blue, short-dashed) the  nonadiabatic case in the zero-range potential. In (f) (green, dot-dashed) only the nonadiabatic momentum shift at the tunnel exit is taken into account in the otherwise quasi-static case of a zero-range potential and experimental data of \cite{Boge_2013} are displayed as black dots.}
      \label{momentum}
    \end{center}
  \end{figure}

We generalize the static picture of tunneling into the nonadiabatic regime as follows. In the quasi-static case, see Fig. \ref{picturelp} (ionization at the peak of the laser field is considered, $t_e=0$, $p_x=0$), the electron tunnels through the potential $V_{\it{eff}}(x) =V^{(z)}-x E_0$ (solid green line) on a constant energy-level (long-dashed black line)  ${\cal E}={\cal E}_\parallel+V_{\it{eff}}$, where ${\cal E}_\parallel=p_x(t)^2/2=A(t)^2/2=E_0^2t^2/2$  
is the kinetic energy along the tunneling direction (negative during the under-the-barrier motion).
In the nonadiabatic case the time dependence of the tunneling barrier should be accurately taken into account: 
${\cal E}_\parallel=A(t)^2/2=[(E_0/\omega)\sin(\omega t)]^2/2$. 
The coordinate dependence of the kinetic energy ${\cal E}_\parallel (x)$ as well as of the energy level at the tunneling  
\begin{eqnarray}
{\cal E} (x) ={\cal E}_\parallel (x)+V_{\it{eff}},
\end{eqnarray}
can be derived (short-dashed blue line in Fig.~\ref{picturelp}) taking into account the electron trajectory under the barrier  $x=\int^{t}_{t_s}dt'A_x(t')$, where the saddle time $t_s$ is determined from the SPM-condition via a $\gamma$-expansion and reads:  
\begin{eqnarray}
 -i\omega t_s&=& \gamma \sqrt{1 + \frac{p_{\perp e}^2}{\kappa^2}} - \gamma^2
  \epsilon\frac{p_{\perp e}}{2\kappa} \sqrt{1 + \frac{p_{\perp e}^2}{\kappa^2}} \label{ts}\\
&& +  \frac{\gamma^3}{24} \sqrt{1 + \frac{p_{\perp e}^2}{\kappa^2}}\left[-4 + 3 \epsilon^2 + 
  4 \left(-1 + 3 \epsilon^2\right) \frac{p_{\perp e}^2}{\kappa^2}\right]+o(\gamma^4), \nonumber
\end{eqnarray}
with $t_e=0$, and the electron transverse momentum $p_{\perp e}$ at the tunnel exit. In a linearly polarized field the coordinate dependence of the energy level can be calculated analytically: ${\cal E}(x)=\kappa ^2 \left[1-\left(\sqrt{\gamma ^2+1}-\gamma  \kappa  x/2 n\right)^2\right]/2 \gamma ^2-x\text{E}_0$, with $n=I_p/\omega$.
In this way one can represent the strong field ionization in the low-frequency regime ($\omega \ll I_p,U_p$), as tunneling with a coordinate dependent energy  which is due to the electron energy gain from the varying barrier. The given intuitive picture suggests that the coordinate of the tunnel exit 
$x_e=\int^{t_e}_{t_s}dtA_x(t)$ shifts  towards the atomic core $x_e=x_{e,qs}-\delta x$ due to nonadiabatic effects
$  \delta x=\left(1-\frac{4\epsilon^2}{9} \right) \frac{\gamma^2}{4}x_{e,qs}$,
where $x_{e,qs}=\kappa^2/2E_0$ is the exit coordinate in the quasi-static case and $\gamma\lesssim 1$. 
As Fig.~\ref{momentum}(b) shows, the coordinate of the electron appearance in the continuum in the nonadiabatic regime is smaller in comparison to the quasi-static case, however, increases with larger $\gamma$, e.g. $x_e\approx 15$ a.u. at $\gamma =4$.

The intuitive picture of Fig.~\ref{picturelp} not only indicates the change of the tunneling exit due to nonadiabatic effects but also can hint how the ionization probability is modified. The tunneling probability in Eq.~(\ref{Gamma}) can be represented via the time-dependent WKB-approximation for $\gamma\lesssim 1$ as follows (for a linearly polarized field) 
\begin{eqnarray}
\Gamma\sim  \exp\left[2i\left(\int_{x_i}^{x_e} p_x (t(x)) dx-\int^{t_e}_{t_s}\left[\frac{p_x(t)^2}{2}-xE(t)\right]dt+I_p t_s\right)\right].\nonumber
\end{eqnarray} In the static, pure tunneling case the last two terms of the equation cancel due to energy conservation, whereas in the pure multiphoton regime of large $\gamma$ the last term dominates and gives the well known $\Gamma\sim {\cal I}^n$-rule, with  the laser intensity ${\cal I}$ and $n=I_p/\omega$. In the intermediate regime that is considered here, all three terms contribute. For $\gamma\lesssim 1$ a modified tunneling exponent can be derived 
\begin{eqnarray}
\Gamma\sim  \exp\left[2i(1+\gamma^2/5) \int_{x_i}^{x_e} p_x (t(x)) dx\right].
\label{Gamma3}
\end{eqnarray}
where the leading order correction due to the last two terms is included. 
According to Eq.~(\ref{Gamma3}) the area between the potential barrier and the energy level in Fig.~\ref{picturelp} can qualitatively indicate the ionization probability. 
Note also that the ionization rate in the $\gamma\lesssim 1$ region  can be approximately factorized as \begin{eqnarray}\Gamma\sim {\cal I}^{n^*}\exp\left[-\int_{x_i}^{x_e} p_{qs}(x)dx\right],\label{Gamma4}\end{eqnarray}with $n^*=\delta {\cal E}/\omega$, the energy change during the under-the-barrier motion $\delta {\cal E}=(x_{e,qs}-x_e)E_0$ and $p_{qs}(x)=\sqrt{-xE_0-(-I_p+n^*\omega)}$, which can be interpreted as $n^*$-photon absorption followed by static tunneling with higher energy ${\cal E}=-I_p+n^*\omega$ (see the red path in Fig. 1 (a)).

As the nonadiabatic corrections lift the energy level up, the tunnel exit shifts closer to the atomic core which increases the ionization probability, displayed as a shrinking of the mentioned area. 
Further, in the case of circular polarization, part of the energy of the tunneling electron is transferred into the transversal direction (see below), decreasing the longitudinal energy ${\cal E}_\parallel$ and, consequently, the energy level ${\cal E} (x)$ which yields a smaller tunneling probability compared to the linear polarization case.
We can  also deduce from Eq. (\ref{Gamma3}) the most probable momentum of the electron at the tunnel exit which corresponds to the minimum of ${\rm Im} \{S(t_e,t_s,\mathbf{p})\}$.
%We approximate the most probable momentum of the electron at the tunnel exit.  It follows from Eq.~(\ref{ts}) that 
In contrast to the quasi-static tunneling case ($\gamma \ll 1$), the tunneling probability $\Gamma (p_{\bot e})$ at an intermediate $\gamma\sim 1$ has a maximum at a non-zero value of the transverse momentum
$  p_{\bot e}=\epsilon\gamma \kappa/6 $, 
see Fig.~\ref{momentum} (c). An order of magnitude estimation confirms that the nonvanishing momentum at the tunnel exit is due to the nonvanishing transversal electric force of the rotating field of the elliptically polarized laser field $E_\perp(\tau_k)\sim \epsilon E_0\gamma$. In fact, the transversal force induces the momentum change
$\Delta p_{\perp}\sim E_\perp(\tau_k)\tau_k\sim \epsilon \gamma\kappa$, with the Keldysh time $\tau_k=\gamma/\omega=\kappa/E_0$. For the most probable tunneling electron trajectory,  this momentum change has to be compensated by a transversal momentum in opposite direction at the ionic core, yielding  at the tunnel exit the electron with transverse momentum in the direction of the transverse force. It is very similar to the relativistic tunnel ionization  where the transversal Lorentz-force is due to the magnetic field \cite{Klaiber_2013c,Klaiber_2013d}. The momentum shift due to nonadiabatic effects is also visible in the asymptotic momentum distribution at the detector. In the static model the maximal final momentum 
is $p_{f}\sim\epsilon E_0/\omega$, while in the nonadiabatic regime the momentum shift during tunneling is added, yielding $p_{f}\sim\epsilon E_0/\omega +p_{\perp e}$, see Fig.~\ref{momentum} (a).

While the simple case of a short range atomic potential was suitable to describe the qualitative modification of the tunneling picture in the case of nonadiabatic ionization,  the effect of the Coulomb field of the atomic core should be taken into account for quantitative predictions \cite{Popruzhenko_2008a,Popruzhenko_2008b,Yan_2012}.
In the case of the  Coulomb atomic potential $V^{(C)}(r)=-Z/r$, where $Z$ is the charge of the ionized atom, the ionized wave-packet in the remote future can be given via
\begin{eqnarray}
  \psi(\mathbf{p}) \sim\int^\infty_{-\infty}dt \int d^3\mathbf{r}\exp[-i S^{(LC)}(\mathbf{r},t)+\frac{Z}{\kappa} \log[x]+i\kappa^2t-\kappa r],\nonumber\\
\label{wfC}
\end{eqnarray}
where the second term in the exponent arises from the SFA-matrix-element and the bound state wave-function~\cite{Klaiber_2013b,Popov_2004u}.
$S^{(LC)}$ is the classical action in the laser and the Coulomb field and fulfills the Hamilton-Jacobi-equation:
\begin{eqnarray}
  -\partial_t{S}^{(LC)}=\left(\boldsymbol{\nabla}S^{(LC)}\right)^2/2+V^{(C)}(r)+\mathbf{r}\cdot\mathbf{E}(t)
\label{WKB}
\end{eqnarray}
The 4-dimensional integral in Eq.~(\ref{wfC}) 
can  be solved with the saddle point method that yields the saddle point conditions for the initial time and coordinate of  the ionizing electron
\begin{eqnarray}
  \partial_t{S}^{(LC)} =\kappa^2/2, \,\,\,\,\,\,\,\,\,\,
  \partial_x S^{(LC)}=i \left(\kappa ^2 \sqrt{x^2}-Z\right)/(\kappa  x),\nonumber
\end{eqnarray} 
assuming that the transversal motion is a perturbation.
Inserting these equations into Eq. (\ref{WKB}) gives $-x \text{E}(t)- Z^2/(2 \kappa ^2 x^2)=0$, 
and the saddle point for the coordinate can be expressed with the time saddle point: $x_s= \exp[-i\pi/3][Z^2/(2\kappa^2 \text{E}(t_s))]^{1/3}$.
From the latter the initial coordinate $x_s(t_s)$ as well as the initial velocity $\dot{x}_s(t_s)=\partial_xS^{(LC)}$ are determined as a function of the initial time $t_s$, with complex values for the saddle time and the coordinate.
The under-the-barrier electron trajectory within the imaginary time method is found by solving Newton's-equation in the Coulomb and the laser field. For the most probable trajectory the coordinate  
becomes real at the tunnel exit ${\rm Im} \{x(t_e)\}=0$. The tunnel exit is defined via $\dot{x}(t_e)=0$.
With these boundary conditions the Coulomb-corrected exit $x(t_e)$ and the transversal exit momentum $\dot{y}(t_e)$ can then be deduced from the solutions of the differential equation, see Fig \ref{momentum}~(b) and (c). 
The reduction of the exit coordinate compared to the zero-range potential case can be understood via the  attractive longitudinal Coulomb force which decreases the tunneling distance. Transversally the Coulomb and the laser force have opposite signs and compensate each other such that the required initial momentum  to come back to the real axis is less for the Coulomb potential than for the zero-range potential case. In total, the under-the-barrier trajectory is more focused along the main tunneling direction due to the Coulomb force of the atomic core.
The initial (saddle point) time is complex with a real part that is negative and of the order of 30 as (see Fig.~\ref{momentum}~(d,e)). 
The time when the electron leaves the barrier, i.e. the tunnel exit time, when the longitudinal velocity is zero, is shown in Fig.~\ref{momentum}~(d,e). It is visible that the electron starts its continuum motion approximately 10 as after the laser field maximum, i.e., in the considered regime
there is a non-negligible time delay in the case of the Coulomb potential due to nonadiabatic dynamics which
vanishes in the limit of small $\gamma$ ($E_0/E_a\ll 1$ is applied for the SPM validity).

Now we turn to the question of the attoclock calibration. For this purpose one has to take into account accurately the Coulomb focusing effect during the electron propagation in continuum after exiting the ionization barrier which affects the photoelectron emission angle.
We use the calculated nonadiabatic time delay as well as the exit coordinate $x(t_e)$ (that is shifted closer to the core due to nonadiabatic effects) and the transverse momentum shift at the exit as starting conditions for the continuum motion. In Fig.~\ref{momentum}~(f) we show that with this simple model experimental data on the asymptotic emission angle can be reproduced if nonadiabatic corrections and the Coulomb field for the under-the-barrier motion are accounted for.  
The quality of this approximation is reduced for large $\gamma$ (small field strength) which explains the deviation from experimental data for these parameters.  

We note that for large field strengths a negative Wigner time delay can be  estimated of an order of magnitude of $\tau_w\sim (E_a/E_0)^{2/3}$ when going beyond the quasi-classical approximation \cite{Klaiber_2013d}. For $E_0=0.1$ a.u. a negative Wigner time delay of approximately 10 as is obtained. This time delay to some extent cancels the non-adabiatic time delay and explains the deviation of the calculated curve from the experimental data in the high field region.

In conclusion, an intuitive model for the intermediate regime of tunneling and multiphoton ionization has been developed. This way the simpleman model of Ref. \cite{Boge_2013} has been adapted to explain its data mostly by a displacement of the tunnel exit.

%in lines of the theory in Ref.\cite{Torlina_2014}  to explain the ionized electron momentum distribution  in the experiment \cite{Boge_2013} with nonadiabatic and Coulomb focusing effects without invoking Wigner tunneling time at not very large laser fields.

%we have developed an intuitive model for the nonadiabatic tunneling ionization which allows to predict in a simple way the modification of the tunneling parameters. We have quantified with this model the role of nonadiabatic effects: a nonadiabatic time delay, a shift of the tunnel exit coordinate and a transversal momentum shift of the electron at the tunnel exit on the attoclock calibration. While these effects partly compensate each other in the following Coulomb focused continuum motion, however, the effect of the shift of the tunneling exit plays a dominant role inducing a decisive impact on the Coulomb focusing  and on the calibration of the attoclock.

MK acknowledges fruitful discussions with Anton W\"ollert, Enderalp Yakaboylu and John Briggs. We also thank Robert Boge and Ursula Keller for providing the experimental data in Fig. 2(f).

\bibliography{strong_fields_bibliography}

\end{document}